# Seasonal Mass Transfer on the Nucleus of Comet 67P/Chuyumov-Gerasimenko

H. U. Keller[1,2*], S. Mottola[1], S. F. Hviid[1], J. Agarwal[3], E. Kührt[1], Y. Skorov[2], K. Otto[1], J.-B. Vincent[1], N. Oklay[1], S. E. Schröder[1], B. Davidsson[4], M. Pajola[5,6], X. Shi[3], D. Bodewits[7], I. Toth[8], F. Preusker[1], F. Scholten[1], H. Sierks[3], C. Barbieri[9], P. Lamy[10], R. Rodrigo[11], D. Koschny[12], H. Rickman[13], M. F. A'Hearn[7†], M. A. Barucci[14], J.-L. Bertaux[15], I. Bertini[16], G. Cremonese[17], V. Da Deppo[18], S. Debei[19], M. De Cecco[20], J. Deller[3], [1]S. Fornasier[10], M. Fulle[21], O. Groussin[10], P. J. Gutiérrez[22], C. Güttler[3], M. Hofmann[3], W.-H. Ip[23], L. Jorda[10], J. Knollenberg[1], J. R. Kramm[3], M. Küppers[24], L.-M. Lara[22], M. Lazzarin[9], J. J. Lopez Moreno[22], F. Marzari[9], G. Naletto[25], C. Tubiana[3], N. Thomas[26]

[1] Deutsches Zentrum für Luft- und Raumfahrt (DLR), Institut für Planetenforschung, Asteroiden und Kometen, Rutherfordstraße 2, 12489 Berlin, Germany

[2] Institut für Geophysik und extraterrestrische Physik, TU Braunschweig, 38106, Braunschweig, Germany

[3] Max-Planck-Institut für Sonnensystemforschung, Justus-von-Liebig-Weg, 3, 37077, Göttingen, Germany

[4] Department of Physics and Astronomy, Uppsala University, 75120 Uppsala, Sweden

[5] NASA Ames Research Center, Moffett Field, CA 94035, USA

[6] Centro di Ateneo di Studi ed Attivitá Spaziali, "Guiseppe Colombo" (CISAS), University of Padova

[7] Department of Astronomy, University of Maryland, College Park, MD, 20742-2421, USA

[8] MTA CSFK Konkoly Observatory, Budapest, Konkoly Thege M. ut 15-17, H1121, Hungary

[9] University of Padova, Department of Physics and Astronomy, Vicolo dell'Osservatorio 3, 35122 Padova, Italy

[10] Aix Marseille Université, CNRS, LAM (Laboratoire d'Astrophysique de Marseille) UMR 7326, 13388, Marseille, France

[11] International Space Science Institute, Hallerstraße 6, 3012 Bern, Switzerland and Centro de Astrobiología, CSIC-INTA, 28850 Torrejón de Ardoz, Madrid, Spain

[12] Department of Physics and Astronomy, Uppsala University, Box 516, SE-75120 Uppsala, Sweden and PAS Space Research Center, Bartycka 18A, PL-00716 Warszawa, Poland

[13] Scientific Support Office, European Space Agency, 2201, Noordwijk, The Netherlands

[14] LESIA, Obs. de Paris, CNRS, Univ Paris 06, Univ. Paris-Diderot, 5 place J. Janssen, 92195, Meudon, France

[15] LATMOS, CNRS/UVSQ/IPSL, 11 boulevard d'Alembert, 78280, Guyancourt, France

[16] Centro di Ateneo di Studi ed Attivitá Spaziali, "Guiseppe Colombo" (CISAS), University of Padova

[17] INAF - Osservatorio Astronomico, Via Tiepolo 11, 34014 Trieste, Italy

[18] CNR-IFN UOS Padova LUXOR, Via Trasea, 7, 35131 Padova, Italy

[19] Department of Mechanical Engineering – University of Padova, via Venezia 1, 35131 Padova, Italy

[20] UNITN, Universitá di Trento, Via Mesiano, 77, 38100 Trento, Italy

[21] INAF – Osservatorio Astronomico, Via Tiepolo 11, 34014 Trieste, Italy

[22] Instituto de Astrofísica de Andalucía (CSIC), c/ Glorieta de la Astronomía s/n, 18008 Granada, Spain

[23] National Central University, Graduate Institute of Astronomy, 300 Chung-Da Rd, Chung-Li 32054, Taiwan

[24] Scientific Support Office, European Space Astronomy Centre/ESA, P.O. Box 78, 28691 Villanueva de la Canada, Madrid, Spain

[25] University of Padova, Department of Information Engineering, Via Gradenigo 6/B, 35131 Padova, Italy

[26] Physikalisches Institut, Sidlerstrasse 5, University of Bern, 3012 Bern, Switzerland

[*]corresponding author e-mail: keller@linmpi.mpg.de

---

[†] Departed on 29 May 2017

**ABSTRACT**

We collect observational evidence that supports the scheme of mass transfer on the nucleus of comet 67P/Churyumov-Gerasimenko. The obliquity of the rotation axis of 67P causes strong seasonal variations. During perihelion the southern hemisphere is four times more active than the north. Northern territories are widely covered by granular material that indicates back fall originating from the active south. Decimetre sized chunks contain water ice and their trajectories are influenced by an anti-solar force instigated by sublimation. OSIRIS observations suggest that up to 20 % of the particles directly return to the nucleus surface taking several hours of travel time. The back fall covered northern areas are active if illuminated but produce mainly water vapour. The decimetre chunks from the nucleus surface are too small to contain more volatile compounds such as $CO_2$ or CO. This causes a north-south dichotomy of the composition measurements in the coma. Active particles are trapped in the gravitational minimum of *Hapi* during northern winter. They are "shock frozen" and only reactivated when the comet approaches the sun after its aphelion passage. The insolation of the big cavity is enhanced by self-heating, i. e. reflection and IR radiation from the walls. This, together with the pristinity of the active back fall, explains the early observed activity of the *Hapi* region. *Sobek* may be a role model for the consolidated bottom of *Hapi*. Mass transfer in the case of 67P strongly influences the evolution of the nucleus and the interpretation of coma measurements.

## 1 INTRODUCTION

Shortly after the OSIRIS camera (Keller et al. 2007) on board the comet rendezvous mission Rosetta resolved the nucleus of comet 67P/Churyumov-Gerasimenko during approach in 2014 its peculiar shape became evident (Sierks et al. 2015). The bi-lobate nucleus rotates with an obliquity of 52° (RA = 69° and DEC = 64°), such that the southern solstice coincides with its perihelion passage, and a pre-perihelion period of 12 h 40 m (Mottola et al. 2014). The size of the nucleus is about 2x4x3 $km^3$ with a volume of 18.8 $km^3$ (Jorda et al. 2016). The mass of the nucleus was determined to 1. $10^{13}$ kg (Pätzold et al. 2016) resulting in a low density body of 533 kg $m^{-3}$. During the early phase of the rendezvous with 67P starting in August 2014, the northern hemisphere was illuminated by the sun and hence the south was not observable. The total extent of the surface could only be determined later in the mission to 47 $km^2$. Early observations of the coma established water to be the most abundant volatile (Gulkis et al. 2015) suggesting that water ice sublimation was the main driver for the cometary activity even at the heliocentric distance > 3.5 au. Once a high resolution shape model SHAP4 (of course with major deficiencies in the south (Preusker et al. 2015)) was available, Keller et al. (2015a) calculated the insolation and ensuing erosion of the surface areas based on a two layer (a thin inert dust layer on top of the cometary dust/ice matrix) thermophysical model. Three cases were considered: i) model A - dirty ice without a desiccated dust layer, ii) model B - (our standard) with a dust layer thickness of 50 µm, 5 times the pore size, and iii) model C - with a dust layer of 1 mm thickness, corresponding to 10 pore sizes.

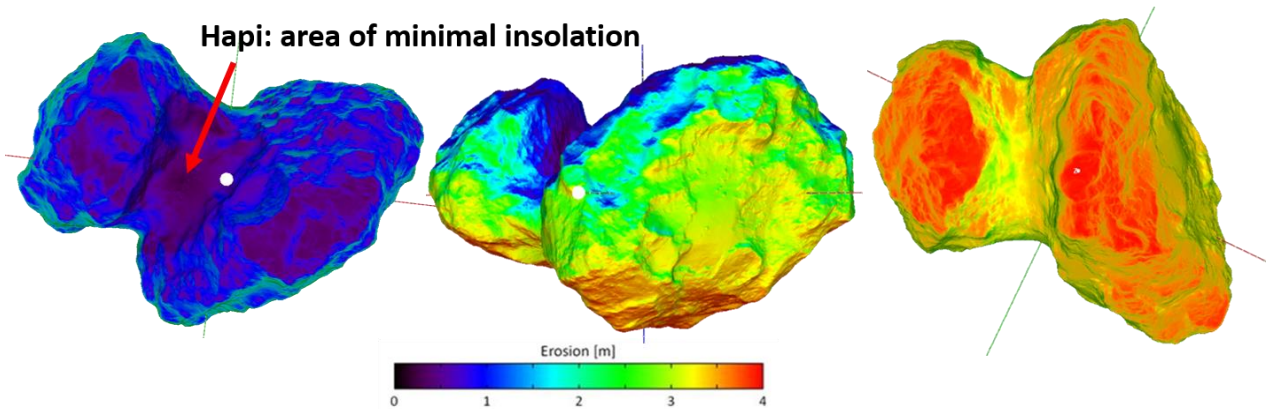

Figure 1 *Water ice sublimation based on the two layer thermodynamic model B (Keller et al. 2015a)*

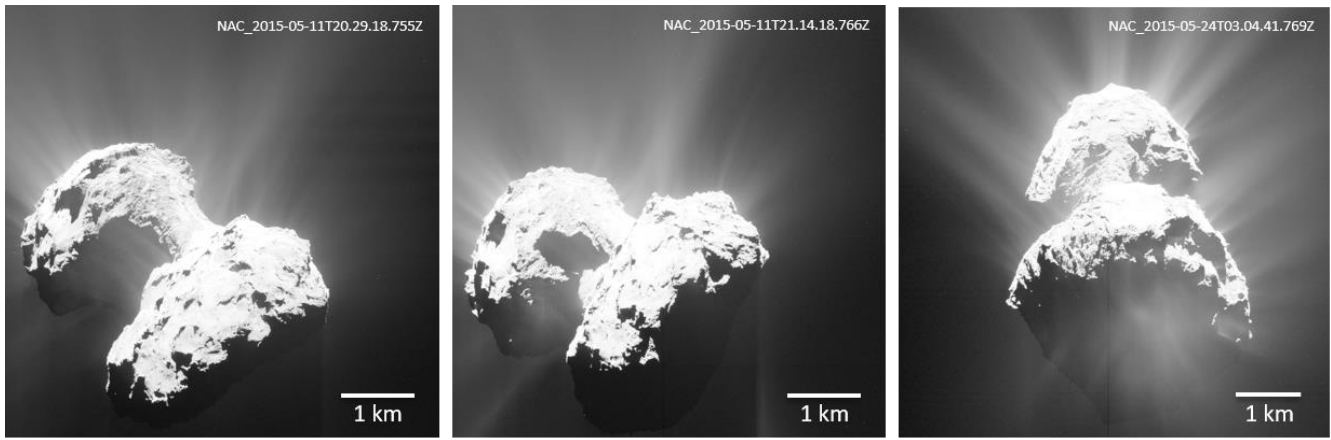

Figure 2 *The cometary nucleus is active wherever the sun shines. Left and middle image taken on 11 May 2015, right image on 24 May 2015. Images are logarithmically scaled and and a gamma corrction is adjusted to enhance the contrast of the weak dust features.*

The current orbit of 67P and the orientation of its rotation axis leads to large differences of the insolation between the north and south hemispheres. Southern summer coincides with the perihelion passage and lasts only 8 months out of the 6.4 y orbital period while the comet travels from the pre-perihelion equinox at $R_h = 1.6$ au to $R_h = 1.8$ au post perihelion. Typically erosion due to sublimation of water ice is four times stronger on the southern hemisphere than on the northern areas (*Fig. 1*). As a consequence Keller et al. (2015a) predicted that the nucleus morphology will show distinct differences accumulated over the last 8 apparitions since comet 67P's orbit changed following a close Jupiter encounter in 1959 (Krolikowska 2003). Based on the same thermophysical model, that assumes the activity to be distributed over a homogeneous surface, Keller et al. (2015b) calculated the variation of the rotation period caused by activity-induced torques. In contrast, the orientation (obliquity) of the rotation axis is hardly changed during the present perihelion passage in spite of the complex bi-lobate shape of the nucleus (Jorda et al. 2016). We can therefore assume that the seasons of insolation have not changed since the last Jupiter encounter. The strong insolation around southern summer solstice could erode up to 20 m during one perihelion passage. Direct determination of the regional loss rates, however, are difficult in the south because during perihelion the resolution of OSIRIS was not sufficient. An overall loss of more than 100 m at specific favourable spots can be expected over the last 6 decades while 67P has been in the inner solar system.

Already during the beginning of the rendezvous of the Rosetta spacecraft it became apparent that the north facing regions such as *Ash*, *Seth*, *Ma'at*, and *Hapi* (for designation of the morphological regions on 67P see Thomas et al. (2015b)) were covered by what looked like a layer of dust. Here "dust" refers to particles with dimensions below the resolution limit of the observing camera. The best OSIRIS observations from distances below 10 km above the surface provided a scale of less than 20 cm per pixel and showed some granularity. Equatorial regions such as *Anubis*, *Anuket*, *Bastet*, and peripheral outer parts of *Imhotep* did not show this cover but rather ragged and coarser terrains. These early observations and the strong dichotomy in erosion between the north and the south suggested that the dust cover on the north results from transport of particles from the south during the southern summer (Keller et al. 2015a). Observations during southern summer confirmed a remarkable morpho-logical dichotomy between the regions on the southern hemisphere and their northern counterparts. Large scale smooth terrains cover most of the north. These and unambiguous depressions are absent on the south that mostly shows consolidated material (El-Maarry et al. 2016). At first it was assumed that the dust cover was inert and jet-like activity would mainly originate from steep cliff surfaces (Vincent et al. 2016). When the comet came closer towards the sun it became clear that all surfaces can be active if they are well insolated (see Fig. 2). In other words, the daily and seasonal activity is controlled by the local insolation. Only recently was it realized (Keller et al. 2016) that the strong contrast in erosion and back fall coverage also leads to a dichotomy of the composition of the coma gases. Indeed, the areas covered by back fall are not inert but, instead, they release water, yet not more volatile species such as CO and $CO_2$.

We will discuss the process of back fall and argue that mass is transferred from the southern hemisphere to the north in the following section. We will provide circumstantial evidence from nucleus observations in Section 3, followed by a discussion of consequences for coma composition observations and their interpretation. Finally we will elaborate on the consequences of the mass transfer for the evolution of the nucleus of comet 67P and discuss the impact on interpretation of former comet observations.

## 2 DISTRIBUTION OF BACK FALL AND SEASONAL ACTIVITY

The distribution of the smooth northern areas was mapped and discussed by Thomas et al. (2015a) who already suggested that these regions are covered by material that fell back onto the consolidated surface. Outcrops and exposure of consolidated material are frequently visible. This indicates that the thickness of the back fall varies from a few decimetres to meters. The actual thickness is difficult to assess but it can be estimated from the thickness of the layer seen at drop offs on top of cliffs (Fig. 3). Other estimates from the shape of a small crater provide a range 1 - 5 m (Thomas et al. 2015a). A layer thickness of ~ 1 m was derived from the observation of seasonal changes (see Sect. 0). The morphology of the imprint of the touchdown of the lander Philae in *Agilkia* suggests the thickness of the dust layer to be 20 cm (Biele et al. 2015).

The peculiar shape of the nucleus forms a large cavity (concave area) between the lobes facing north located not far from the north pole. This area also corresponds to the gravitational minimum on the nucleus and is therefore a preferred location for the accumulation of back falling material. This "neck" area (called *Hapi*) does not display consolidated material with the exception of a few relatively big blocks that have been originally conceived to be boulders that tumbled down from the cliffs of both sides of *Hapi*. These blocks are aligned in the centre of *Hapi* along a ridge connecting more consolidated flat hill tops on both sides (Penasa et al. 2017) and we therefore now interpret them as outcrops. The barely visible ridge creates a small maximum in the geopotential (see Fig. 4 centre image and Fig. 5). Comparing the south oriented part of the neck *Sobek* to the smooth *Hapi* region reveals another striking example of the dichotomy. *Sobek* is not covered with back fall material in contrast to *Hapi* although it constitutes the second strongest gravitational poten tial minimum (Fig. 4). The terrain of *Sobek* appears chaotic with large pinnacles (several tens of metres high) and is framed by cliffs on the small lobe (*Neith*) and on the body lobe (*Geb* and *Anhur*), similarly steep but not quite as high as the outstanding *Hathor* wall on the north side of the small lobe (Fig. 5). Most images of *Sobek* were taken nadir looking and it is difficult to perceive the roughness of the valley bottom and steepness of the walls. A slanted view of an image co-registered on the shape model provides a more realistic impression (Fig. 6). It is conceivable that the consolidated ground of *Hapi* before it was filled with back fall, looked similar to *Sobek* today. The blocks along the center of the valley would then represent the tops of pinnacles. This suggests that the thickness of the deposit is several tens of meters. Lai et al. (2017) found a net thickness increase of 0.4 m at *Hapi* during the Rosetta mission considering only particles < 45 mm. The present accumulation would then take 50 - 100 apparitions of 67P on its present orbit. More probable is that *Hapi* is filled by large chunks that could move mass more efficiently into the gravitational sink. If the filling started with the last orbit change of 67P, a depth increase by 3 m per orbit would yield a total of 24 m. This scenario is reinforced by the consideration that big blocks observed in flight (Fulle et al. 2016) are so fragile that they often will not survive impact, particularly not on a low geopotential region. This is corroborated by the massive collapse of the *Aswan* cliff (Pajola et al. 2016b) that left only small (0.5 to 10 m) blocks in the talus. The impact speed of chunks falling back is several times higher (in the extreme even higher than the escape speed of ca. 0.8 m s$^{-1}$) than what results from a crumbling cliff wall with a height of 100 m (free fall results in 0.2 m s$^{-1}$). Indeed, the images of the back fall covered areas all show granular material below the resolution limit of a few decimetres but only a few blocks (compare also Fig. 7). Big pieces will readily break up during impact due to their minute tensile strength, required for separation from the surface by the low gas drag forces (Skorov et al. 2017) and/or by interior cracks. Back fall from the north onto the south is less and easily removed by the onset of activity on the south. A secondary gravitational minimum lies in the centre of *Imhotep* near the equator and is also deeply covered by back fall material. Here dramatic changes were observed shortly before perihelion (Groussin et al. 2015) leading to progressive depressions (scarps) of several meters height. This suggests that the covering material is several or even many meters thick. The local gravitational potential is important for the collection of back fall material (Fig. 4).

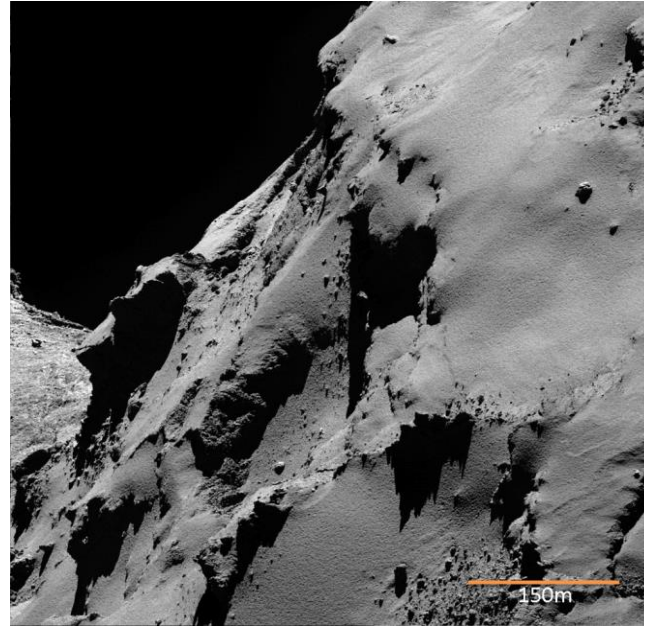

*Figure 3* Back fall covering *Seth*, mostly showing a smooth granular surface texture. Near cliffs the back fall seems to slip. NAC image NAC_2016-03-12T16.03.34.539Z_IS20_1397549300_F22, scale 0.32 m/px.

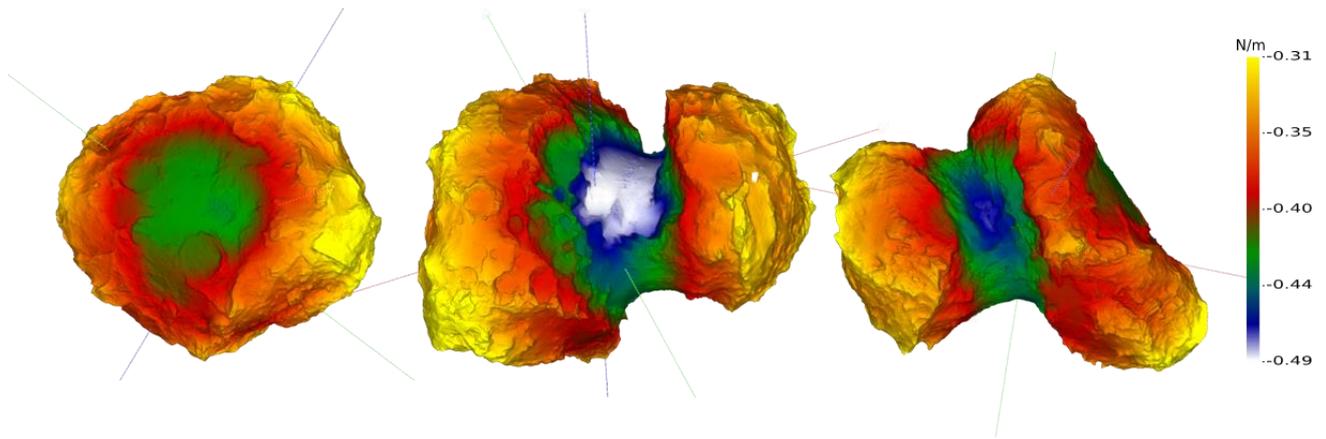

Figure 4 The gravitational potential of the nucleus of 67P. Left Imhotep, middle Hapi, and right Sobek. The geopotential minimum at Hapi shows a small relative maximum along the center indicating a ridge.

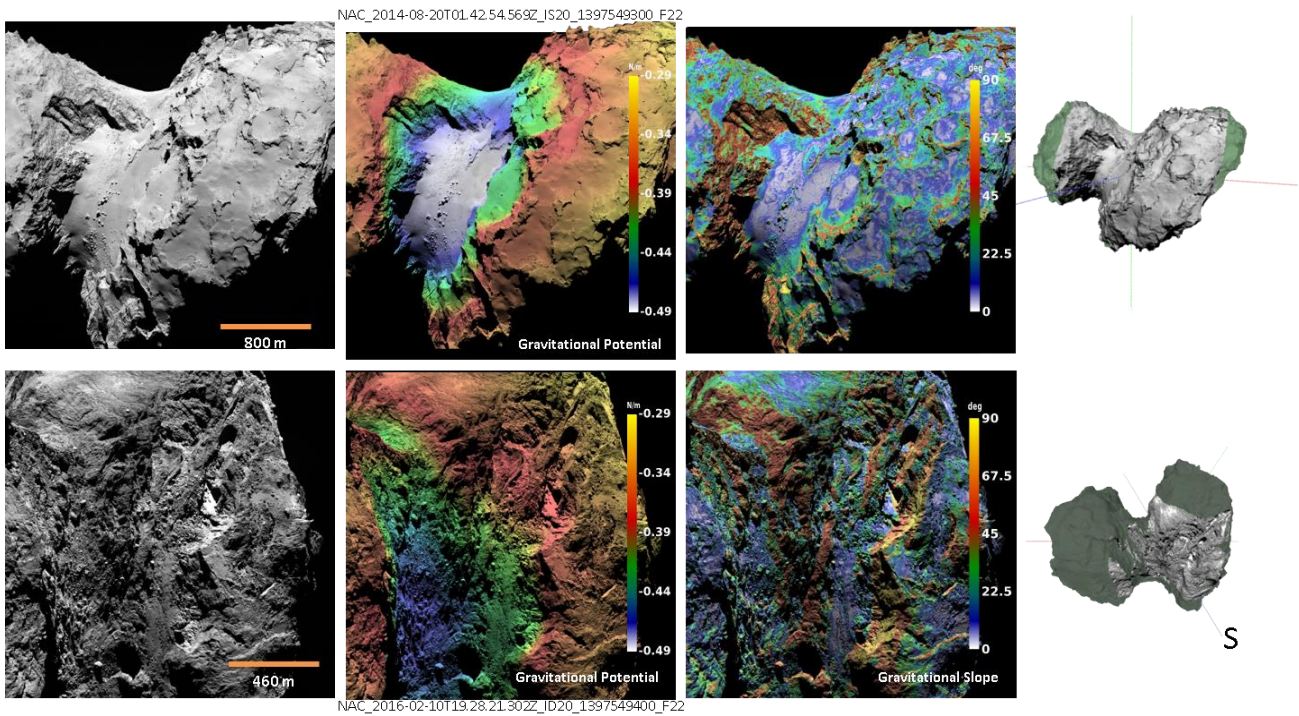

Figure 5 A comparison of Hapi and Sobek. Upper row shows an image of *Hapi*, the geopotential, and gravitational slopes. Lower row an image of Sobek, the geopotential, and gravitational slopes.

*Wosret* on the south side of the small lobe receives some of the most intense insolation integrated over the cometary orbit, 40 times more than what reaches *Hapi*. The morphology of consolidated material is rough. Decimetre sized boulders (comparable to the resolution limit) are strewn mainly in the frequent grooves (Fig. 8). The marked transition from the highly active south-facing region *Wosret* to its neighbouring

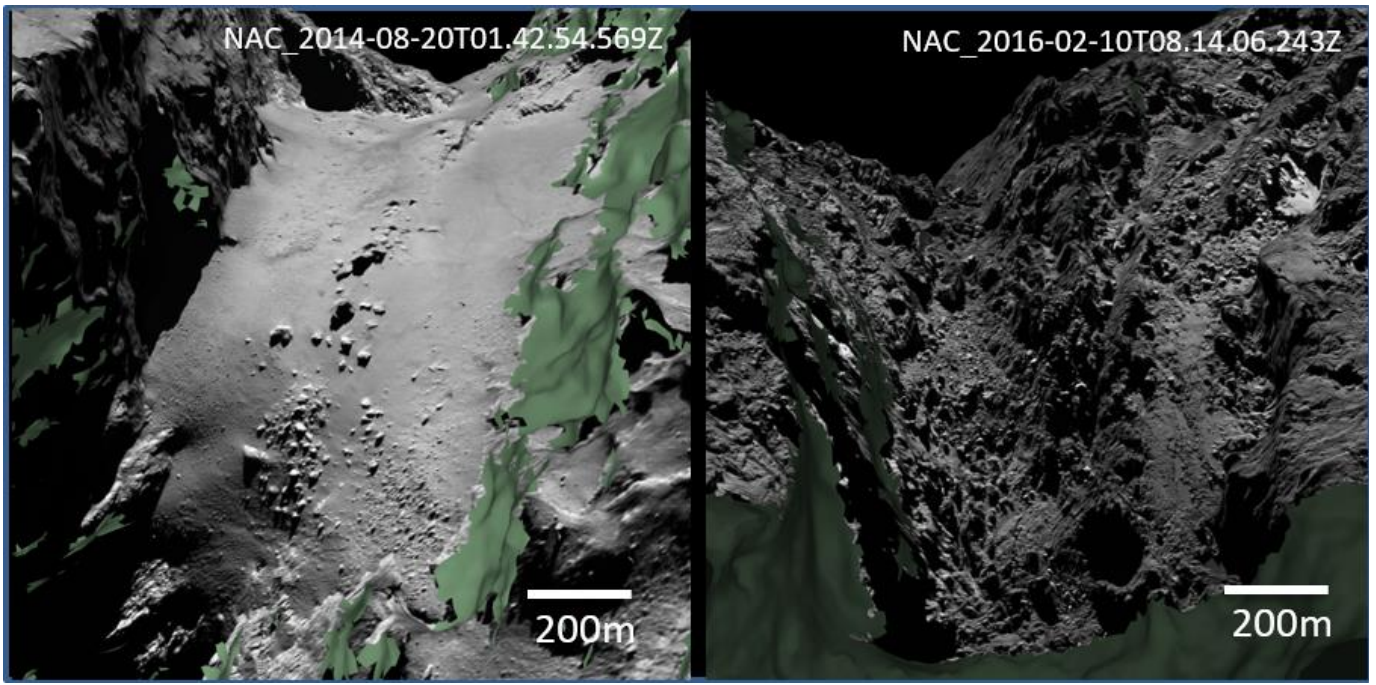

*Figure 6* Comparison of *Hapi* and *Sobek* in slanted views of the images of *Fig. 5* co-registered on the shape model of the nucleus of 67P. Parts of the shape not covered by the images appear green. Outcrops mark the underlying ridge in the middle of the *Hapi* valley. The ground floor of *Sobek* is extremely rough with long pillars, not found in this concentration anywhere else on the nucleus (Basilevsky et al. 2017)

north-facing *Maftet* is striking evidence of the dichotomy (Fig. 9).

High resolution observations by the lander camera ROLIS just before the first touchdown of Philae (Mottola et al. 2015) revealed wind tail like features around obstacles (boulders) in *Agilkia* (Fig. 10). These are oriented in south-north direction caused by impinging particles coming from the south. Similar features were found on OSIRIS images further north from *Agilkia* (see Fig. 10). This implies a global mass transfer from the south to the north. Thomas et al. (2015a) invoked directed particle saltation to explain dune-like forms and wind tails in the *Hapi* area. Jia, Andreotti & Claudin (2017) suggested that lateral coma winds at the day/night boundary can move decimetre sized particles within a turbulent zone just above the surface to form the dune-like features.

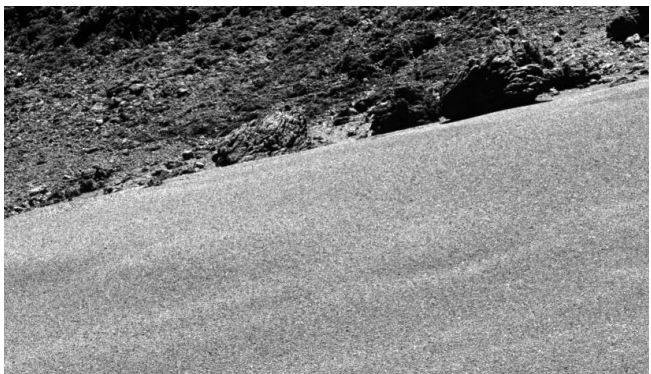

Figure 7 Slanted view towards the north rim of *Imphotep*. The back fall cover appears very smooth with a granularity close to the resolution of the image. No boulders are found. NAC_2016-05-19T16.06.08.961Z_IS20_1397549300_F22 resolution 15 cm/px at the rim. The width of the left boulder is *35 m*

Many OSIRIS images taken above the limb show dust particles leaving the active nucleus surface. In order to be detectable on OSIRIS images the size of the lifted particles has to be larger than millimetres (see Fig. 6 of Thomas et al. (2015a)). At the time of perihelion, when Rosetta was far from the nucleus, the size of large particles could be determined using the spacecraft motion to provide a stereo baseline (Fulle et al. 2016), thereby covering a large volume. The ejected mass was concentrated in chunks of 10 cm and bigger. The cumulative mass loss rate of particles with < 1 kg (r = 8 cm) is only 1/10 of the total rate. In the vicinity of the spacecraft the stereo effect between the wide and narrow angle cameras of OSIRIS can be exploited (Ott et al. 2017) yielding good agreement with the mass loss rate determined by Fulle et al. (2016). The biggest flying particle (boulder) resolved by OSIRIS passed the Rosetta spacecraft at a distance of 3.4 km on 30 July 2015, 180 km from the nucleus. Its dimensions were estimated to be 1 m times 0.5 m. Systematic observations of particles within the Hill sphere of the nucleus in August 2014 at a heliocentric distance of about 3.5 au revealed about 350 particles on bound orbits around the nucleus (Rotundi et al. 2015). Their sizes were estimated to lie between 4 cm and 2 m. Particles above about 5 cm can survive on bound orbits through the aphelion passage of the comet (Richter & Keller 1995). Models show that the fraction of particles in bound orbits is about 0.1 % of the total dust mass (Fulle et al. 1997). Very specific geometric conditions and/or perturbations need to be fulfilled for particles to enter bound orbits. Most of the large particles will follow ballistic trajectories to be widely distributed over the surface or escape into the cometary trail (Agarwal et al. 2010). The largest particle ejected at the time of observation by GIADA had a diameter of 17 mm when the activity level was still very low at ca. $R_h$ = 3.5 au. The dust mass loss was strongly concentrated in large particles. All these lines of evidence suggest that almost all of the dust mass is lost in large particles of decimetre and bigger size.

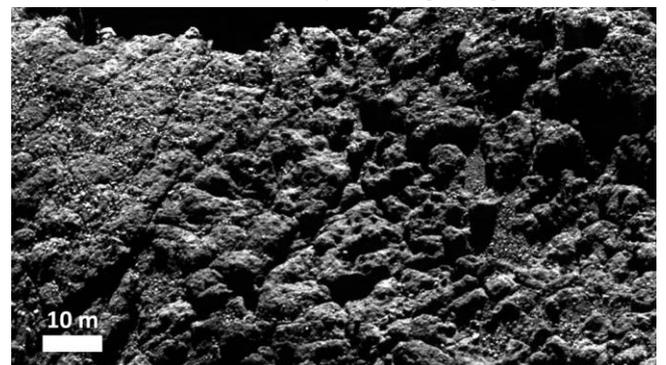

Figure 8 *Consolidated material in the centre of Wosret with leftover boulders near the resolution limit of 11 cm/px. NAC_2016-05-19T11.08.37.677Z_IS20_1397549000_F22 cropped.*

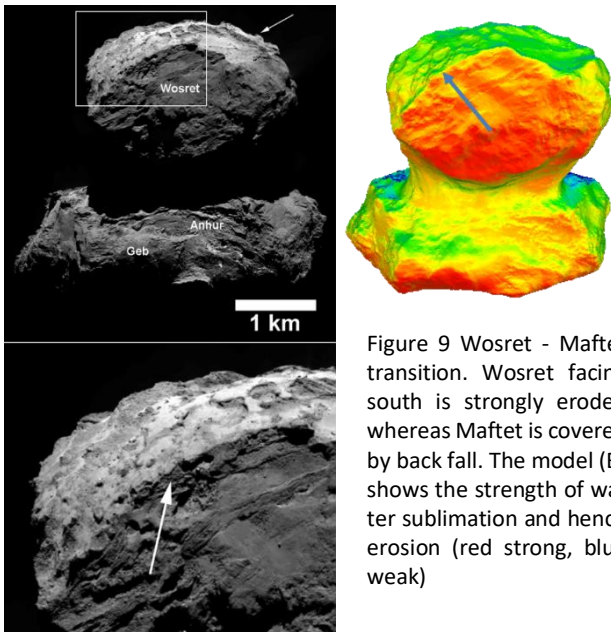

Figure 9 Wosret - Maftet transition. Wosret facing south is strongly eroded whereas Maftet is covered by back fall. The model (B) shows the strength of water sublimation and hence erosion (red strong, blue weak)

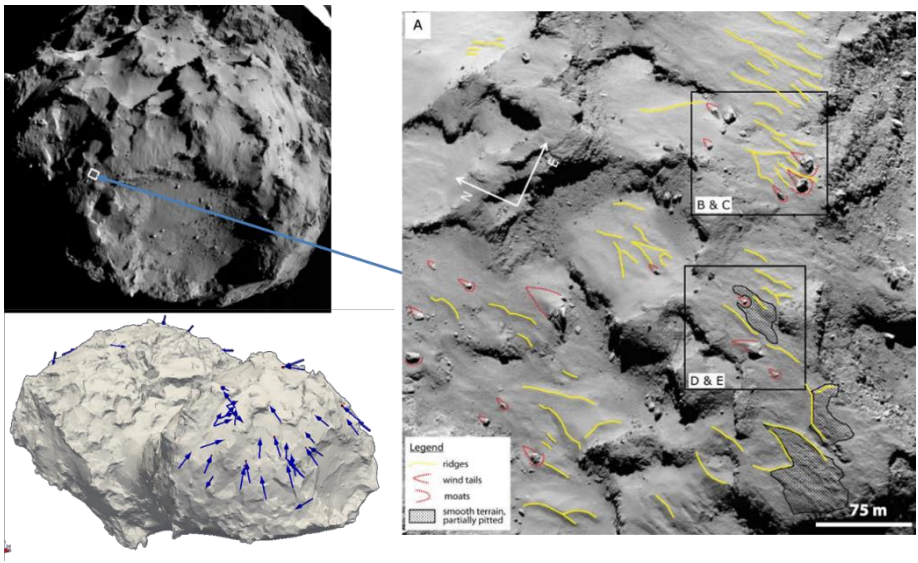

Figure 10 Wind tails around obstacles observed by ROLIS at *Agilkia* (Mottola et al. 2015) and registered over the northern hemisphere (Otto et al. 2017). Blue arrows indicate the directions of wind tail features.

All observations of smooth dust-covered surfaces show granularity down to the resolution limit. Images of the *Imhotep* area taken after perihelion show an extremely uniform granular surface indicating structures in the 10 to 20 cm range (Fig. 7). Even higher resolution images (< 2 cm/px) taken just before the spacecraft crashed in *Ma'at* (Pajola et al. 2017b) and by ROLIS (0.95 cm/px) show granular material inter dispersed with larger chunks (Fig. 11 and Fig. 12). The ROLIS images show two different surface units with slightly different particle size distribution functions. Pajola et al. (2016a) suggest that the somewhat rough unit is covered by a thin layer of back fall material.

In addition to the strong water sublimation on the south, more volatile compounds such as $CO_2$ contribute and may be required to release large dust particles and chunks. Lai et al. (2017) investigated the redistribution of small dust particles. They combined a simplified thermophysical model of the seasonal water sublimation from a homogeneous nucleus surface with DSMC calculations of the gas drag taking also into account the gravitational field but not radiation pressure that is of importance for small particles. They considered dust particles in the size range from $3\ 10^{-3}$ to 45 mm (Fulle et al. 2016). Over the whole perihelion passage of 67P the *Hapi* area showed the strongest gain, being covered by almost 0.4 m of dust. A further concentration of dust was found in the central *Imhotep* plane. Here is the gravitational low not as distinctive and in addition the loss of dust at the equator is higher than at *Hapi*. The concavity formed between the lobes makes *Hapi* the least illuminated area of the comet. Implicitly, Lai et al. (2017) assumed that the dust back fall did not quench the activity. Realistic models of the back fall distribution would first of all require the understanding of the physical process of lifting dust particles by the subliming gases. The distribution of deposits will be strongly influenced by the nature and local variability of the activity. This includes the strength of the activity that provides the gas drag to lift and accelerate the particles, the complex gravitational field, the rotation of the nucleus, the radiation pressure for small sub millimetre particles, and any reactive force coming from outgassing large particles themselves. Kramer & Noack (2015, 2016) calculated trajectories for a uniformly active nucleus including gas drag, gravity, and Coriolis force. They found agreement with the orientation of the wind tails observed by ROLIS (Mottola et al. 2015).

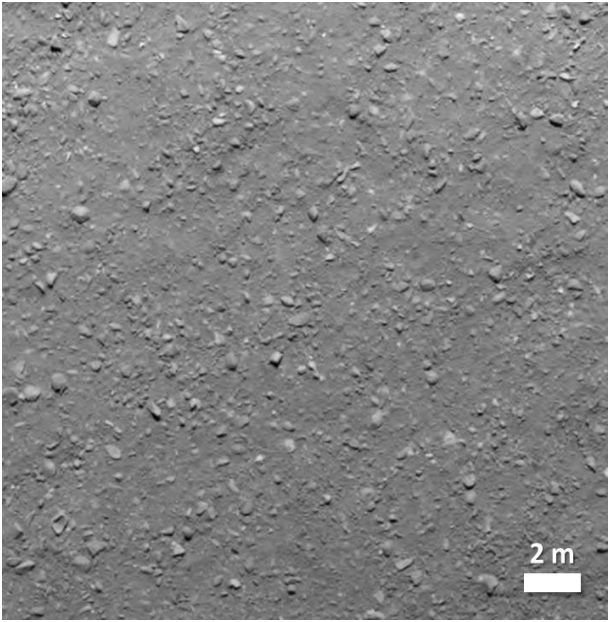

Figure 11 High resolution image of back the fall covered area *Sais* shortly before the crash of the Rosetta spacecraft. WAC_2016-09-30T10.32.05.769Z_IS20_1397549000_F12, resolution 3 cm/px, size 672 x 672 px

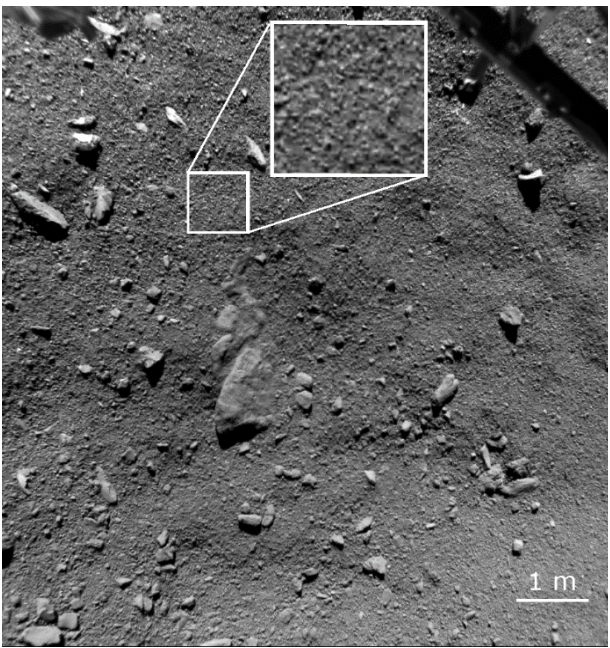

Figure 12 ROLIS image at *Agilkia* (Mottola et al. 2015).

## 3 EVIDENCE FOR ACTIVE BACK FALL

What fraction of particles leave the near-surface gas acceleration zone with velocities below escape (ca. 1 m/s)? For a rough estimate of the amount of deposit (back fall) that can be expected from the south we assume that all the eroded mass leaves the surface in large chunks (Fulle et al. 2016) that can be observed by OSIRIS..

Specific observational sequences taking images in various cadences and exposure times while the NAC was staring at the limb of the nucleus could be used to follow trajectories of particles ejected from the surface (Agarwal et al. 2016). 76 particles were identified to leave the nucleus in almost radial direction. Assuming similar photometric properties as derived for the nucleus surface the size of the identified particles peaked at a radius of 10 cm. For smaller sizes, the sampling was incomplete due to background particles. Towards larger radii the differential size frequency distribution decreased with an exponent of -4. If we assume that these aggregates are part of the nucleus material, a 10 cm particle has a mass of 2.2 kg if the mean density of the nucleus of 533 kg m$^{-3}$ is applied (Pätzold et al. 2016). This mass coincides very well with the peak of the particle mass distribution determined by Fulle et al. (2016).

The average velocity in the image plane of the particles was comparable to the local escape speed around 0.6 m s$^{-1}$. The particles were subject to constant accelerations over the 2 h of observations. Surprisingly, 50 % of the aggregates were accelerated towards the nucleus, away from the sun. The lateral acceleration components were equally distributed. One particle could be pursued over a long trajectory arc. Its motion changed direction becoming finally directed towards the nucleus (in anti-solar direction). If the particles (as a broken away part of the nucleus) contain water ice then its sublimation creates a rocket effect (Tauber & Kuhrt 1987). The effective acceleration will mostly be directed in anti-solar direction even if the aggregate rotates. Kelley et al. (2013) investigated similar particles in the coma of comet 103P/Hartley 2. The strength of the rocket effect should be comparable to the initial acceleration by the gas drag assuming water vapour is responsible for the liberation of the particle. As long as the particle remains active, the rocket effect will be constant while the radial gas drag decreases with distance from the nucleus due to the decrease of the gas density. About 10 % of the observed particles had a speed in excess of the local escape velocity and will most probably leave the nucleus. The fate of the particles is certainly difficult to predict.

In an effort to provide some insight, we calculate the trajectories of the icy chunks observed on 6 Jan. 2016 and described in Agarwal et al. (2016). We assume that they were subject to gravity, $a_{grav}$, of the big lobe and a constant acceleration characteristic of each chunk individually, $a_{const}$, such that $a_{grav} + a_{const} = a_{mean}$, where the latter is the acceleration measured from the trajectories in the image plane, generally $< 2\ 10^{-4}$ m s$^{-2}$ about twice the gravitational acceleration. The trajectories are given by $\ddot{x}^i(t) = a^i_{const} + a^i_{grav}(x,y)$, $\dot{x}^i(t = t_0) = v^i(t_0)$, $x^i(t = t_0) = x_0^i$, where the index i refers to the image x and y directions, respectively, $x_0^i$, $v_0^i$ are the projected positions and velocities at the beginning of the observation sequence, $t_0$, and $a^i_{grav} = - GM\, x^i / r^3$, with r being the projected distance from the centre of mass. We approximate the big lobe by a sphere with a radius of 2 km and the corresponding mass, $6.6 \times 10^{12}$ kg (Jorda et al. 2016). In this model, 10 % of the chunks fall back to the nucleus (sphere). The time to intersection with the surface is on the order of 1 - 6 h, with a mean of 3.6 h, corresponding to ~1/4 of a comet rotation. The chunks from this particular active area in *Khonsu* would therefore land on the surface within *Imhotep* in the southern hemisphere.

To estimate an upper limit (under the present model assumptions) for the fraction of chunks falling back, we also integrated the trajectories setting the constant acceleration to zero for times beyond the end of the observation series (2 hours after $t_0$), such that the chunks were then only subject to gravity. In this case, 18% of the chunks would fall back to the surface with flight times between 1 and 27 hours, of which most are relatively equally distributed between 1 and 12 hours. The material would be distributed across the whole nucleus. We note that the above model ignores the motion in the direction along the line of sight, for which we do not know the initial position, velocity or acceleration. We find that the number of chunks with negative vertical acceleration that escape from the nucleus gravity field is twice that of returning particles. The escape of these aggregates is mainly due to the additional horizontal acceleration. The calculations suggest that the observed acceleration towards the nucleus is not sufficient for aggregates to hit the nucleus. More global 3-D models using the observed initial conditions will provide a more realistic estimate of the back fall.

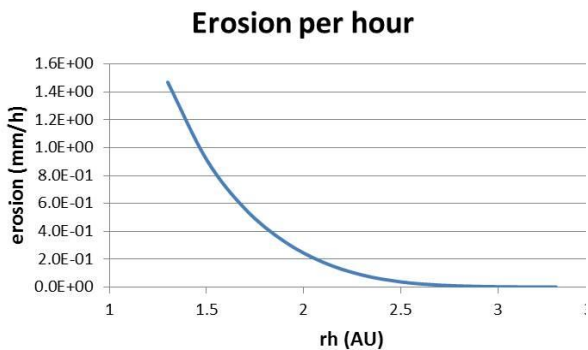

Figure 13 Erosion rate of a fast rotating sphere as function of the heliocentric distance.

The transit times of the particles falling back near their origin on direct trajectories are only a few hours. For this reason, they lose little mass during the flight. Assuming a fast rotating sphere (in all directions) of dark ice/dust mixture (no dust cover) with an albedo of 0.05 yields an upper limit for the mass loss. In addition we assume a density of 533 kg m$^{-3}$, a dust/ice mass ratio of 5, and a low heat conductivity of 0.001 W/mK. The aggregate radius shrinks about 1.4 mm per hour at perihelion (Fig. 13). Particles arriving on more distant regions will take considerably longer but large decimetre sized particles can easily survive several 12 h rotation periods of 67P.

If aggregates descent towards the nucleus above active surfaces they may not reach the surface because they are again accelerated by the local gas drag. Particles on originally short trajectories may undergo several hops until they reach an area of little or no activity. The fact that back fall particles do not land on active regions can be well studied at the border between *Wosret* and *Maftet* (Fig. 9) on the top of the small lobe. *Wosret*, near the southpole, is active during the perihelion passage whereas *Maftet* is less favourably illuminated. The surface of *Maftet* is covered by the smooth dust layer while *Wosret* is free of dust. The transition is clear and sudden.

Near horizontal planes of most of the north-facing regions such as *Ash*, *Babi*, *Seth*, *Maftet*, *Serqet*, *Hapi* and at least partially *Nut* and *Ma'at* and near the equator, central parts of *Imhotep*, are covered by a smooth layer of back fall material (Thomas et al. 2015a). An estimate without detailed mapping amounts to about $15 \pm 5$ km$^2$, about 1/3 of the nucleus surface. While during the early rendezvous phase the first dust activity was mainly connected to *Hapi* (Sierks et al. 2015) it became obvious that the comet shows activity wherever the sun shines (Fig. 2). All the smooth back fall surfaces are active as well as surfaces showing consolidated material. Detailed investigations of images taken in April 2015, when the comet was at $R_h = 1.8$ au, by Shi et al. (2016) showed that dust activity originates from smooth surfaces in *Ma'at*. Foot points of dust features that become more obvious during sunset are located on smooth regions and are not connected to outcrops, cliffs or holes. We deliberately avoid the term "jet" because the features hardly show any divergence, they rather look like columns with foot prints defining their diameters (see Fig. 2 of Shi et al. (2016)). We interpret these features as small local inhomogeneities in dust density in the otherwise more or less homogenous gas flow from the dusty surface. This unambiguously confirms that dust-covered surfaces are active and activity is not quenched as originally assumed. The *Ma'at* area was observed at high resolution shortly before equinox (at $R_h = 1.78$ au) at the end of northern summer. At many places the dust cover seemed to be removed from the consolidated nucleus and "honeycomb" patterns with typical extents of 50 to 100 m became visible (Fig. 14) (Shi et al. 2016). During the same time interval changes in the *Hapi* region were found too (Davidsson & others 2017). In the central part of *Imhotep*, the dust cover is assumed to be several or even many meters thick. Groussin et al. (2015) observed fast moving morphology changes (scarps) somewhat later in June 2015 just after equinox.

Numerous changes, including the honeycomb features were found in the dust covered areas that were located between 15° and 50° north, moving south with the waning northern summer. (Hu et al. 2017) concluded that the distribution and timeline of changes (loss of dust cover) suggest that the erosion was more likely driven by the sublimation of water ice than that of more volatile species. They estimated the thick-ness of the dust layer by comparison of the morphology before and after the appearance of the honeycomb structures using photoclinometry. A loss of material with varying thickness of up to 1 m was found across a lateral scale of 10 to 20 m. Features observed just before equinox were covered again on images taken a couple of months after perihelion (Fig. 14). This seasonal variation, coverage by material chunks from the south during southern summer around perihelion and depletion during the following northern summer is a result of the strong activity and production of large decimetre sized "wet" chunks that will crumble and being carried away by the activity during northern summer.

physical mixture of water ice with mostly inert organic material are unknown. The bluing is correlated with an increasing level of activity and indicates that the physical properties of the very surface change with insolation and activity. All the cometary surfaces, dusty areas as well as the consolidated material, confirm this change. This again suggests that the whole surface is active if heated sufficiently, in particular also the dusty regions rich in back fall material.

A region that stands out, when compared to other areas, is *Hapi*. Here the first dust (jet) activity was detected. The neck has been consistently bluer than the nearby *Seth* area (Fig.

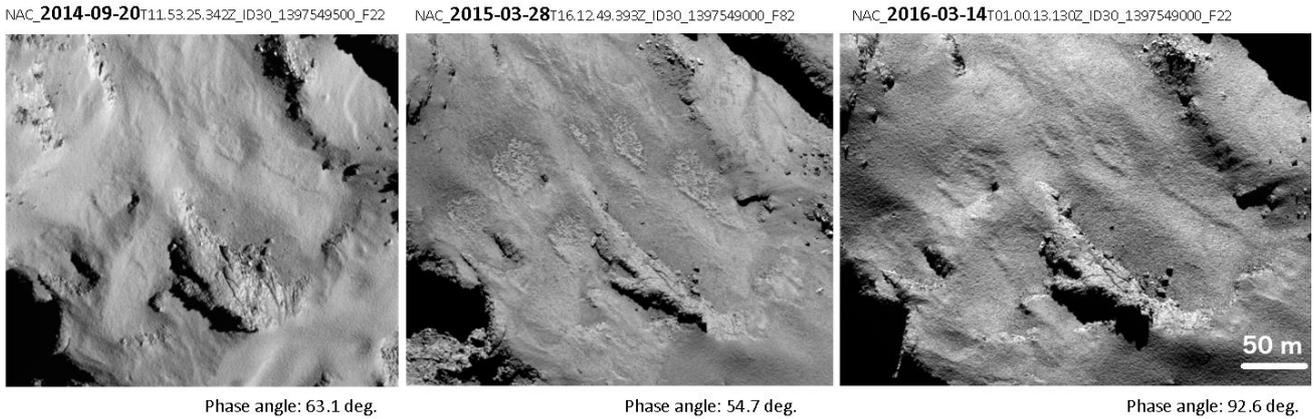

Figure 14 Seasonal changes of honeycomb features in *Ma'at*. The thickness of the seasonal cover is about 1 m (Hu et al. 2017).

Photometric observations of the nucleus showed that the colour of the surface changed with the distance to the sun. The spectral slope, characterized by NAC images taken with filters between 535 and 882 nm (Oklay et al. 2015) shows bluing (a less steep spectral slope) when the comet approached the sun (Fornasier et al. 2015) and reddening again when it

15). Here water was detected following the morning terminator (De Sanctis et al. 2015) indicating that surface material contained water ice that re-condensed during the night on or near the surface. Models interpreting the coma observations by VIRTIS (Bockelée-Morvan et al. 2015, Fougere et al. 2016a), MIRO (Gulkis et al. 2015), ROSINA (Fougere et al. 2016b, Hansen et al. 2016), and OSIRIS (Marschall et al. 2016) identify *Hapi* as particular active for water and dust during the approach to perihelion up to about equinox. During the early phases of the mission, it was argued that the infrared radiation from the wall of *Hathor* on the small lobe and the cliffs of *Seth* on the large lobe enhances the energy input in the neck cavity just above what is available on the convex parts of the northern comet surface (Sierks et al. 2015). However, at that time with 67P at a heliocentric distance around 3.6 au, the energy input onto *Hapi* during one rotation seemed not sufficient and therefore a large compositional or structural difference of *Hapi* was suggested in order

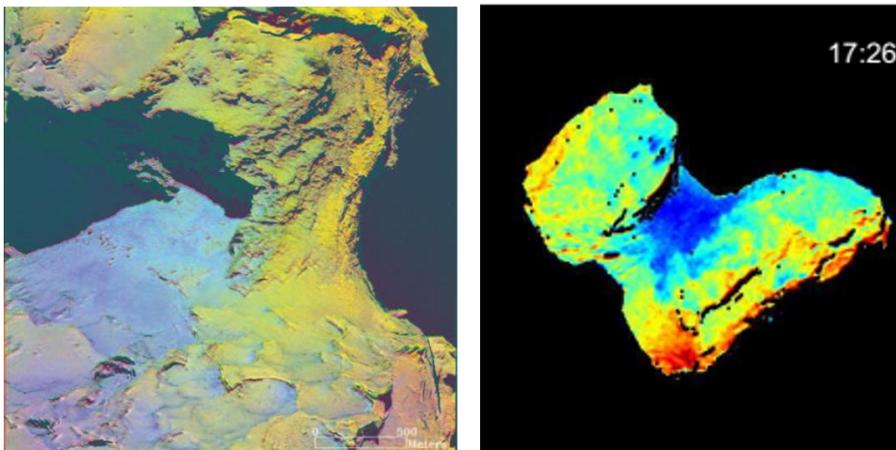

Figure 15 The central parts of *Hapi* in an RGB image are bluer, richer in near surface ice, than the *Seth* area. OSIRIS observations at $R_h$ = 3.5 au on 1 (right, (Fornasier et al. 2015)) and 6 (left, (Oklay et al. 2016)) Aug. 2014 pre-perihelion

receded from perihelion (Fornasier et al. 2016). Water ice patches detected by VIRTIS (Barucci et al. 2016, Filacchione et al. 2016) correlate well with high resolution NAC filter observations showing a blue slope. Hence, this bluing is generally interpreted as the spectral signature of water ice becoming more visible. On the other hand, the albedo of the surface changed very little. The details of the to explain its activity. Averaged over the smooth *Hapi* region the enhancement by self-illumination is only about 50 % while the total insolation reaches only 70 % of well insolated areas in *Seth* (Keller et al. 2015a). So, why is the back fall material at *Hapi* more active? We will discuss this phenomenon in Section 6.

## 4 THE DICHOTOMY OF THE GAS COMA COMPOSITION

The ROSINA mass spectrometer (Balsiger et al. 2007) has a large acceptance angle and hence can resolve areas on the nucleus only when the spacecraft is very close to the surface. Due to the rotation of the nucleus and the motion of the spacecraft, the distance and the illumination of the surface change continuously. Therefore it is difficult to correlate observed species to sources on the surface and to determine their activity level. Early observations of the coma composition comparing the relative abundances of $H_2O$, $CO_2$, and CO showed that observed variations of these compounds were not correlated (Hässig et al. 2015). The $CO_2/H_2O$ density ratio peaked above southern *Imhotep*. This led to speculations about the heterogeneity of the cometary nucleus. Modelling of the gas expansion with the aim of achieving information about the source regions of the species revealed that the very volatile $CO_2$ and CO originated from the equatorial and illuminated southern areas. Fougere et al. (2016b) found a concentration of the water production rate near *Hapi* during the inbound leg of 67P's orbit, whereas the strongest $CO_2$ concentration occurred above *Khonsu* near the southern part of *Imhotep*. The northern areas covered with back fall outgassed only small amounts of the highly volatile compounds. Imaging the column densities of $H_2O$ and $CO_2$ by VIRTIS clearly showed the disparate origin of these species (Bockelée-Morvan et al. 2015, Fink et al. 2016). $H_2O$ concentrated above the northern areas and $CO_2$ above the illuminated equatorial and southern regions. The ratio of the production rates of $CO_2$ and $H_2O$ decreased on approach to the sun, as expected from the difference of volatility.

Earth-based spectroscopic observations of comet 67P showed a strong asymmetry of CN, a daughter radical of the volatile HCN molecule. CN could be only detected shortly before perihelion but was present beyond the southern equinox (Snodgrass et al. 2016). Again a hint that highly volatile molecules hardly appear in the gas coma while the north is active.

As discussed earlier, the back fall particles are in the decimetre range and smaller. This is comparable or smaller than the skin depth under which $CO_2$ ice can be expected. The actual thickness of the layer (of dust and water ice) above the $CO_2$ sublimation front depends on pore size, heat conductivity of the grain material, and diffusive permeability of the layer (Gundlach et al. (2015) and Skorov et al. (2017)).For any *bona fide* parameter set the layer above the $CO_2$ sublimation front amounts to several decimetres. The depth of CO is considerably larger because of its very low sublimation temperature and should be at least several meters (Skorov et al. 2016). Even centimetre sized aggregates may still contain water ice if the effective pore size is in the 100 μm or smaller range. The back fall aggregates are therefore "wet" and contain water ice but no or very little, more volatile, species. This explains why back fall covered areas become active when the comet returns from its aphelion and its surface reaches temperatures high enough to sustain water ice sublimation. At the same time the cover of back fall material isolates the underlying cometary nucleus and prevents sublimation of $CO_2$ and reduces the sublimation of CO. This explains the observed dichotomy in the coma composition. Consequently, the measured $H_2O/CO_2$ ratio over northern areas was high during the early rendezvous (Fougere et al. 2016a, Hässig et al. 2015). The ROSINA measurements revealed strong variations of the $H_2O/CO_2$ ratio, $CO_2$ dominating in the south that was barely illuminated. Abundance ratios determined during northern summer do not reflect in any way the composition of the nucleus. This is different for observations of the southern hemisphere around and shortly after perihelion during southern summer. Here only small patches of back fall material exist that have little influence on the overall activity and composition of the coma. Water molecules, however, are somewhat underrepresented. The back fall does not only transport refractory material but also the water ice inside the aggregates. This amount of water is missing in the coma during southern summer. To establish the correct ratio of water to the more volatile species one needs to integrate over the whole orbit. Even then, if some of the back fall will not sublime on the north, the amount of water ice will be underestimated.

## 5 MASS TRANSFER FROM SOUTH TO NORTH

So far, we have very limited insight in the fraction of dust that falls back in the northern hemisphere. The thickness of the lost layer of cometary material in the south depends on the dust to ice ratio and on the density of the nucleus. Model calculations that assume a pore and particle size of a few to tens of micrometre for the dust ice matrix show that several meters up to 20 m of the nucleus can be lost by activity. As long as we do not understand the process of particle lifting, the arguments are based on the energy balance and therefore speculative concerning the details. Nevertheless, a loss of surface material in the order of meters is reasonable and may be confirmed by the analysis of OSIRIS images. A major drawback, however, is that we obviously do not have observations of the southern hemisphere before the onset of activity because it was not illuminated. During peak activity the Rosetta spacecraft was many hundreds of kilometres away from the nucleus and the image resolution lower than the expected loss by activity. The situation is different for the northern hemisphere. Here we have good mapping from the early rendezvous and from the end of the mission when the sun was already again at northern latitudes. The accuracy of the resulting shape models is about one meter (Jorda et al. 2016). The expected changes, however, are also smaller and at best comparable to the accuracy of the shapes.

The estimate based on photoclinometry (Hu et al. 2017) of the honeycomb features suggests a layer thickness in the order of 1 m. The thickness of back fall will of course vary depending on the topography and morphology of the surface, e. g., the gravitational slope. The thermal models suggest that a layer of 1 m can be removed during northern summer. Consequently, more heavily covered areas should accumulate back fall material because the lower activity of the northern hemisphere is unable to remove large amounts of back fall.

Observations by the OSIRIS cameras ((Agarwal et al. 2016) see Sect. 0) suggest that a considerable part of large aggregates fall back on the nucleus surface. If we assume that about ¼ of those finally rests on the northern plains the back fall and its removal would about balance. The covered nucleus in the north is "protected" from erosion by the back fall. Hence the loss of consolidated nucleus material from the north is very limited. All the erosion of the nucleus takes place during southern summer. Consequently the northern hemisphere has seen little changes at least as long as the comet has been on its present orbit (see also Vincent et al. (2017)). The deep pits recall the images of the nucleus of 81P/Wild 2 (Brownlee et al. 2004) and suggests a comet that arrived in the inner solar system only recently. The exceptions are the cliffs that are not covered by back fall and still face towards the sun near and closely after northern equinox. Models (Keller et al. 2015a) clearly reflect this enhanced insolation and suggest that the cliffs progress (see Fig. 1, left image). Activity from cliffs was already observed for comets 19P/Borrelly (Britt et al. 2004) and 9P/Tempel 1 (Farnham et al. 2013).

## 6 SEASONAL COURSE OF ACTIVITY AND EROSION

We calculated the water production of the nucleus along its orbit at 5 different spots distributed over the whole nucleus in the regions of *Seth, Hapi,* and *Ma'at* facing north, *Imhotep* at the equator, and *Wosret* near the south pole (Fig. 16). We assumed a surface with homogeneous physical properties and applied the parameters for the two-layer thermophysical model that correspond to model B of Keller et al. (2015a).

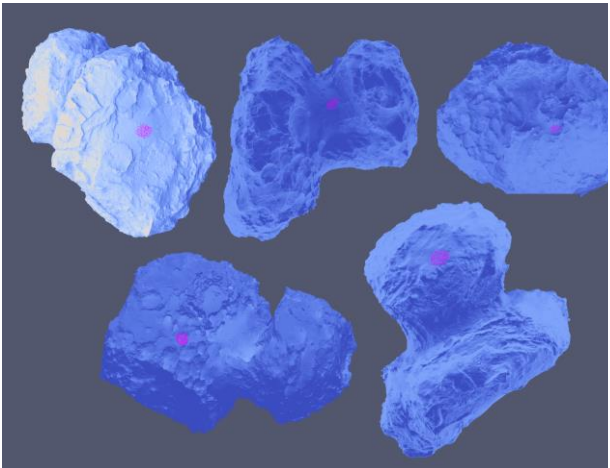

Figure 16 The activity of water sublimation was calculated for 5 regions distributed over the nucleus of 67P. The facets are indicated in pink. Clockwise starting in the left upper corner: *Imhotep*, *Hapi*, *Ma'at*, *Wosret*, and *Seth*.

For all northern regions, the water production rate per unit surface at 4 au is already an order of magnitude higher than at 5 au, increasing rapidly towards perihelion (Fig. 17). Shortly afterwards, around 2.7 au the production rate of *Hapi* reaches its low maximum and decreases from thereon until aphelion. The production rates of *Seth* and *Ma'at* continue to increase culminating near northern equinox at 1.6 au. They then sharply drop to zero just shortly before perihelion. At 1.6 au after perihelion the production of water started again to reach a low maximum at 2.2 au. The production rates of the northern regions follow the insolation that is controlled by the peculiar shape of the nucleus and the obliquity of the rotation axis. *Hapi,* lying in the cavity between the two lobes, is only reached by the sun for relatively short intervals of a cometary day. The shading by the lobes becomes more and more dramatic when the sun approaches equinox.

As it can be expected, the trend for regions that can be reached when the sun is at southern latitudes is dramatically different. The water production for both *Imhotep* and *Wosret* strongly peaks at perihelion following the solar flux variation. The peak of the activity of *Wosret* is considerably narrower, more confined to the highest subsolar latitudes, than that of *Imhotep* that is still well insolated at the equinoxes.

MIRO detected water activity already during the final approach phase of Rosetta on 6 June 2014, when the comet was at 3.92 au from the sun (Gulkis et al. 2015). Observations by ROSINA started at the beginning of August 2014 at a heliocentric distance of 3.6 au revealing a $CO_2$ production rate that was not correlated with water ranging from a few percent to tens of percent (Hässig et al. 2015). Even at 4 au the activity of 67P seemed to be already driven by water sublimation. This onset of water sublimation agrees well with our model.

It is difficult for *in situ* measurements (ROSINA and GIADA) to separate the contributions of the different areas. Coma observations looking tangential to the nucleus limb (VIRTIS) are affected by similar difficulties. In addition to seasonal variability, the daily variability (following the insolation) considerably complicates the issue and makes it almost impossible to resolve regional inhomogeneities if they exist. Only MIRO can detect water when looking down on the nucleus surface, however with a relatively low resolution due to its large field of view.

To assume the same physical parameters for the consolidated or back fall covered areas is justified as long as the pore size is small and the back fall aggregates can be treated as clasts. The observed increase of water production (Hansen et al. 2016) is steeper than what our models predict (Keller et al. 2015a). A way to reconcile this discrepancy may be derived from the colour observations by OSIRIS. The spectral slope of the surface, becomes bluer, when the comet approaches the sun (Fornasier et al. 2015). This suggests that water ice becomes more visible and hence can more easily sublime when the comet approaches the sun. The physical properties of the surface change with activity a development that is not considered in our model. A bluer surface is equivalent to an effectively thinner dust cover, and can be simulated by a transition from model C to model A without dust cover (Keller et al. 2015a).

A big puzzle remains. Why did *Hapi* appear to be the most active area during northern summer although it receives relatively little insolation? The spin-integrated water produc-tion pre-perihelion does not stand out and overall is the lowest of all the areas, with nearly no production after northern equinox (Fig. 18). A hint comes again from the spectral slope observations. *Hapi* is by far the bluest region, right from the beginning of the Rosetta rendezvous at 3.5 au (Fig. 15). *Hapi* shows more water ice on its surface than the plains covered with the same back fall material. What separates *Hapi* from all other regions is that the insolation of *Hapi* around and also after perihelion is practically zero. The rest of the northern regions (*Seth* and *Ma'at*) are insolated enough around the southern equinox that they still show an activity that is comparable to our model calculations for *Hapi* pre-perihelion (Fig. 17). The insolation becomes weaker while the comet recedes from the sun. Around 4 au and beyond, the absorbed energy is not sufficient anymore to produce a gas flux strong enough to lift the particles. The dust remains on top of the dust/ice matrix material and forms a desiccated layer that will have to be removed or thinned once the comet comes back towards the sun. This is fully in accord with observed variation of the spectral slope becoming bluer during approach and turning red after southern equinox (Fornasier et al. 2016).

The secluded location in the big cavity formed between both lobes provides a strong enhancement of up to 50 % additional energy input by self-heating from the *Hathor* wall and the *Seth* slope when the sun is at high northern latitudes during early approach of the comet toward perihelion. Particularly near the border of *Hapi* with *Hathor* the water sublimation is enhanced (Fig. 19). By using our standard model B (dust layer thickness 50 μm corresponding to 5 pore sizes) and integrating over one spin at $R_h$ = 3.93 au we find that the water production in the *Hapi* region is 3 to 4 times higher than what is typical for the dust covered areas *Ash*, *Seth*, and *Ma'at*. The peak activity rate is even higher considering that most of *Hapi* is in shadow of the small lobe over about half the spin period whereas the northern territories are in permanent sun light.

At *Hapi*, the back fall aggregates arrive at a surface that is very cold because it does not see sunlight during northern winter. Surface temperatures measured on the southern hemisphere during its winter time are far below 80 K (Gulkis et al. 2015). The aggregates that are active during their transit as long as they are in sunlight are shock frozen when they enter the shadowed zone. At these low temperatures no water ice sublimes. The water vapour coming from the still warm interior of the particle will form frost near and on its surface driven by the thermal inertia (see Shi et al. (2016) discussing sun set jets). In contrast to other northern areas,

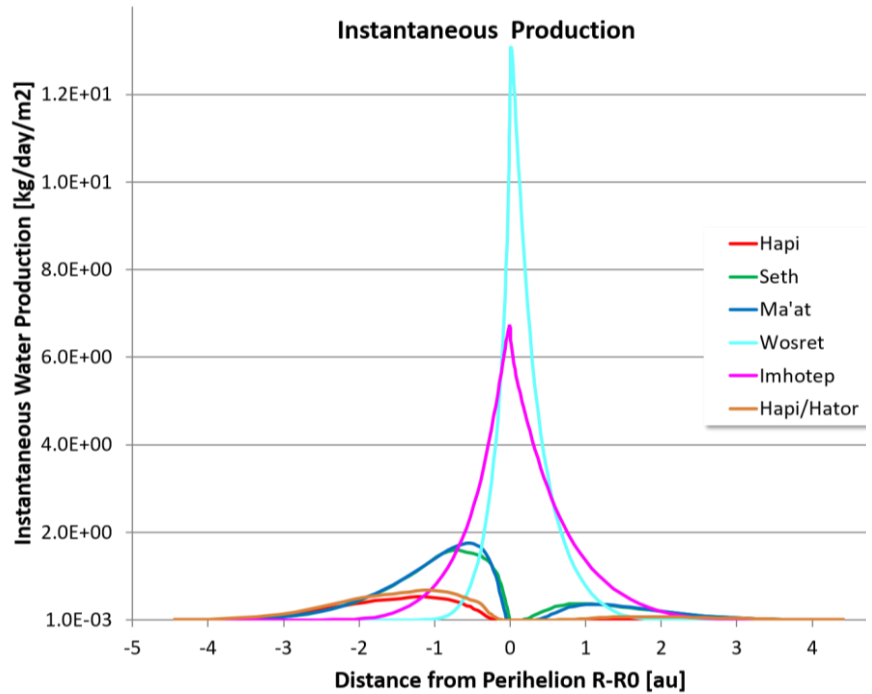

Figure 17 Water production rates integrated over one spin period at the 5 spots indicated in Fig. 16. In addition the water production rate was calculated for a region of *Hapi* bordering to Hathor where the IR self-heating is particularly strong.

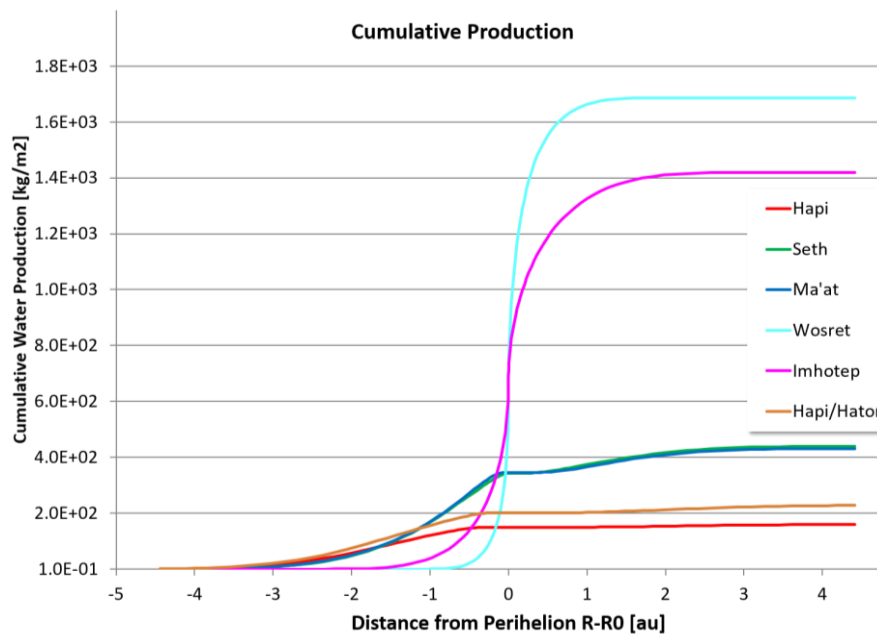

Figure 18 Cumulative water production rates at the 5 spots indicated in Fig. 16. In addition the water production rate was calculated for a region of *Hapi* bordering to *Hathor* where the IR self-heating is particularly strong.

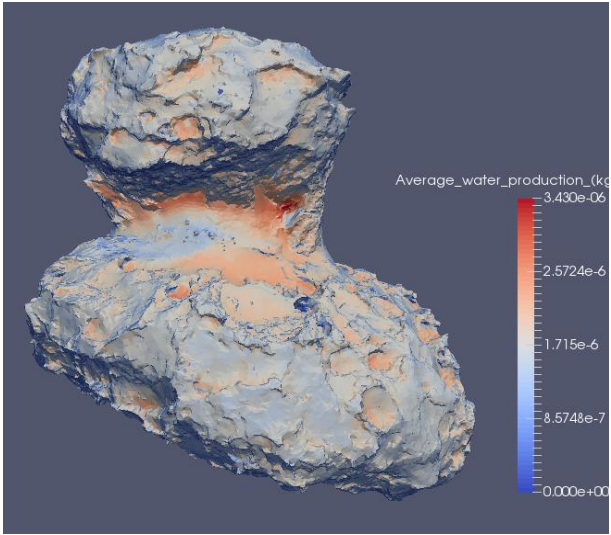

Figure 19 Water activity averaged over one spin of the nucleus of 67P at R$_h$ = 3.93 au showing *Seth* and *Hapi*. Standard model B (Keller et al. 2015a)

local activity in the *Hapi* valley is at least an order of magnitude higher at noon than what is produced at other places of the northern hemisphere. In addition, when observed with the line of sight along the valley of *Hapi*, dust brightness is integrated (Keller et al. 2015a). This easily explains why OSIRIS first detected dust jets here (Sierks et al. 2015). Our calculations and estimates demonstrate that *Hapi* is the strongest local source of water activity during the comet's approach towards the sun. In the scope of our thermodynamic model, the enhancement is due to a thinner desiccated dust cover on *Hapi* than on most of the northern hemisphere. This layer forms during the inactive aphelion passage. Strong self-heating supports the activity even further. No enhancement of water content is required.

During northern summer *Hapi* is shadowed by the small lobe when most of the northern hemisphere is continuously insolated. This modulation should be visible in the water production if *Hapi* is the major contributor. Later, close to the northern equinox the big lobe also casts shadow and the intervals of activity will be even shorter.

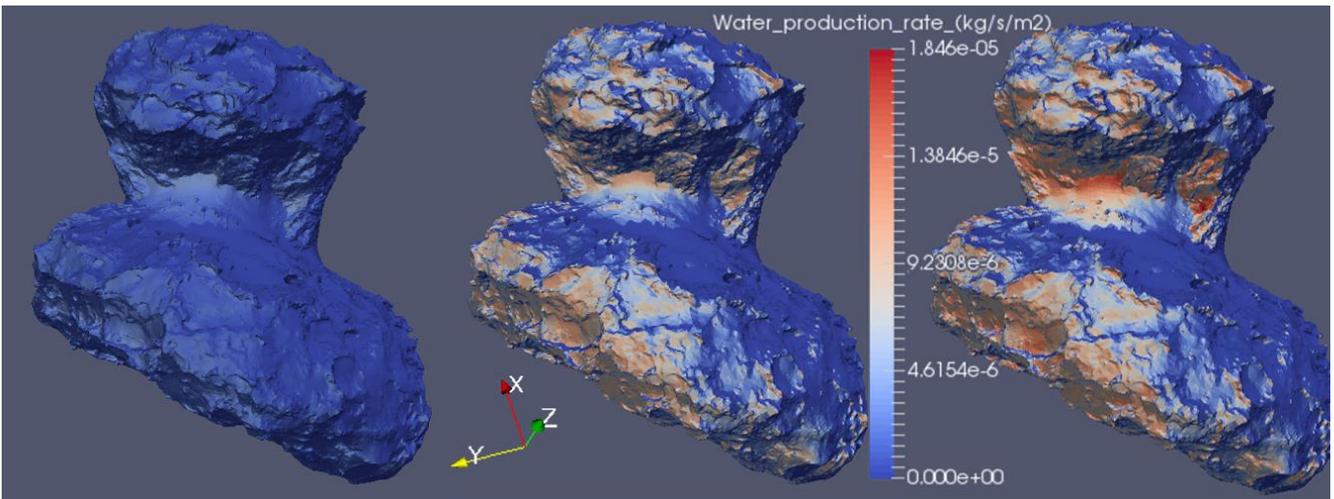

Figure 20 *Instantaneous water production rates at R$_h$ = 3.93 au using model C with self-heating (left), model A without self-heating (middle), and model A with self-heating (right). All views use the same color scale.*

the temperature stays low until the comet emerges from aphelion. A desiccated layer will not build up and hence the spectral slope will remain blue like that of active surfaces already at large heliocentric distances (Fig. 15). The missing desiccated surface layer leads to substantial higher activity than what our standard model B provides. Fig. 20 shows the instantaneous water production calculated for a spin position when the sun shines into the *Hapi* area. We compare the activities calculated with model C, model A without self-heating, and model A with self-heating. The maximum activity in *Hapi* for the last model is about a factor four higher than the maximum of model C. Compared to the activity found on the dust covered plains the activity is enhanced by an order of magnitude. The values of model A with self-heating still underestimate the maximum possible activity. The temperature of the *Hathor* wall should correspond more to model C (196 K with a thin desiccated dust cover, about 15° higher than the bare surface assumed for model A) and provide more self-heating than what model A yields. Therefore

## 7   DISCUSSION AND CONCLUSIONS

OSIRIS observations of the surface at a resolution high enough to derive the size distribution of the back fall are scarce. The ROLIS observations (Mottola et al. 2015) at *Agilkia* and the last images before the Rosetta crash at *Sais* in the *Ma'at* region allow for a detailed analysis of the dust size distribution. Several dust components can be identified. Pajola et al. (2017b) correlate the size distributions with varying exponents to back fall from several regions of the south and north hemispheres with activity levels varying according to the season. They assume that the lift pressure of the subliming ice determines the size of the expelled particles. The relative strong activity of *Hapi* pre-perihelion creates back fall at *Agilkia* and *Sais* both located on the northern side of the small lobe. *Hapi* itself, *Agilkia*, and *Sais* are covered around perihelion by bigger back fall particles originating from *Bes* on the south side. The paper suggests that back fall and re-distribution of back fall is a common phenomenon and highly variable with season.

Finally, changes on the surface of 67P have been observed ubiquitously, at first in back fall regions like *Imhotep* (Groussin et al. 2015) and *Hapi* (Thomas et al. 2015a) (Davidsson & others 2017), but also on *Ma'at* and a vast part of the dust cover (Hu et al. 2017, Shi et al. 2016). The massive collapse of the *Aswan* cliff (Pajola et al. 2017a) is spectacular, whereas most of the other changes are subtle and near the resolution of the images (El-Maarry et al. 2015, 2016). The overall, highly asymmetric, shape of 67P suggests a highly active past. The current distinct dichotomy of the morphology and topography advocates an orbital constellation constant for longer than the last 8 apparitions. The shape models of the nucleus reach their final accuracy (Jorda et al. 2016, Preusker et al. 2015) but it will be difficult to derive global changes because only the northern hemisphere could be systematically imaged pre- and post-perihelion. Here of course the erosion was weaker when compared to the erosion of the south during the perihelion passage. The northern surface was not only eroded but also deposits were found, e. g., on the honeycomb patches. It may eventually be possible to create a difference model of the northern hemisphere. The interpretation of mass loss due to activity and deposition from back fall is difficult to disentangle because processes are dynamic and vary with topography. The analysis of the Rosetta spacecraft trajectories around and near the nucleus of 67P (Pätzold et al. 2016) will finally determine the total mass loss of the comet during the present perihelion passage. A preliminary value is $1.1 \pm 0.3 \, 10^{10}$ kg (Pätzold Rosetta SWT 47) which corresponds to only 0.1 % of the nucleus mass. Averaged over a mass-equivalent sphere with a density of 533 kg m$^{-3}$ only 0.55 m were lost. Using instead the determined surface of the nucleus yields 0.44 m. We can then calculate the losses on the southern and northern hemispheres taking into account the overall sublimation ratio of 4:1. This yields 0.8 m and 0.2 m, respectively, for an average loss of 0.5 m. The breakeven point where material that is eroded on the south and in part transferred to the north during southern summer and then completely eroded during northern summer is a southern loss rate of slightly more than 1 m. This implies that about 20 % of the eroded material in the south is transferred to northern regions. If the back fall is not evenly distributed over half the surface (24 km$^2$) but concentrated in the smooth areas with about 15 km$^2$ the thickness is accordingly bigger and back fall will accumulate. This is currently definitely the case for *Hapi* and large parts of *Imhotep*. The erosion of ca. 1 m is surprisingly small compared to the maximal erosion capability of up to 20 m according to the thermophysical models with a thin desiccated layer on top of the ice/dust matrix of the nucleus (Keller et al. 2015a). This implies that the desiccated dust layer may be thicker quenching the activity more than modelled or that the surface is not evenly active in a sort of checkerboard pattern. A thicker deposit (Hu et al. 2017) requires a larger fraction of back fall which at this time does not seem probable but could be achieved by predominantly big chunks. Nevertheless the balance of erosion and back fall providing an ever smaller loss fraction into the tail and trail becomes more delicate.

We have discussed the observations that support the process of mass transfer from the southern hemisphere to the northern territories.

1) The regions *Ash, Babi, Seth, Maftet, Serqet, Hapi* and at least partially *Nut* and *Ma'at*, about 2/3, of the northern hemisphere and partly *Imhotep* at the equator appear smooth and are covered by granular material..
2) The neck region *Hapi*, the gravitational minimum, is filled with granular material many meters thick. Outcrops from the *Sobek*-like valley bottom provide an estimate of its depth. Rapid changing scarps and dune-like phenomena were observed.
3) A secondary gravitational minimum, *Imhotep*, is also filled with granular material showing even more extensive expansions of meter high scarps
4) Most of the southern hemisphere is free of granular material. *Sobek*, the counterpart of *Hapi*, on the south side of the neck is free of dust due to its high level of activity around perihelion. *Sobek* may be a role model for the bottom of *Hapi* before it was covered by back fall.

Model calculations show that the insolation and erosion potential on the south during perihelion passage are 4 times higher than that on the north.

Most of the mass of the released dust aggregates is contained in decimetre and bigger chunks. These are nucleus material and are wet, containing water ice, but not more volatile compounds like $CO_2$ or $CO$. The observations supporting the activity of the back fall aggregates are:

1) Tracking trajectories of released aggregates requires an acceleration component towards the nucleus that is caused by the repulsive force of sublimation towards the sun direction.
2) All the surface of 67P is active when insolated including the back fall covered areas.
3) The *Hapi* region is particularly active during the early approach of 67P towards perihelion.
4) Activity from the northern hemisphere is strongly biased to water vapour.

The active wet chunks constitute an extended source for water molecules as observed in the coma of comet 103P/Hartley 2 (Kelley et al. 2013). This has so far not been observed during the Rosetta mission and sets an upper limit for the additional sublimation, again suggesting transport by big particles. Mass transfer from the highly active regions to the inactive little or non-illuminated areas is an important part of the activity cycle. In the case of 67P mass transfer strongly influences the evolution of the nucleus. The fraction of material landing on the nucleus rather than being lost into the dust tail and trail is difficult to assess. It is large enough to fill *Hapi* and parts of *Imhotep* with meters of material and cover most of the north facing horizontal areas. In *Hapi* the back fall accumulates with time because the overall energy input in this north facing cavity is not sufficient to completely sublimate it. The situation in *Imhotep* is better balanced. Insolation over the whole orbit should be sufficient to remove meters of material. On the contrary the observations of the moving meter-high scarps suggest a depth of the back fall material of many meters. At the end of northern summer some spots in *Seth* and *Ma'at* became dust free (e. g. the honeycomb features) while others were still covered. Pristine wet back fall preserved in the *Hapi* cavity until the comet returns from aphelion explains the blue spectral appearance and the early activity observed by Rosetta. The bluing of the surface during the approach of the comet towards the sun suggests that the (thin) desiccated dust cover is reduced and again built up after perihelion producing the observed reddening. The cover of the northern territories restrains the erosion of the nucleus proper. The unilateral erosion influences the shape of the nucleus and leads to a strong north-south dichotomy of the morphology and topography.

The connection point between the small and big lobes does not lie along the line connecting the centres of mass. The small lobe seems to tip over towards *Seth*. This can be readily explained by the much stronger erosion on the south side (*Wosret*) over extended times. The north (*Ma'at*) like *Seth* has seen little change and is in this sense more pristine.

The size distribution of the back fall material is dominated by large particles and does not represent the released size distribution. This also influences the observed dust size distribution in the coma. The dust layer, devoid of super volatiles, strongly distorts the observed composition in the coma producing a dichotomy. Highly volatile compounds are underrepresented. Measurements above the southern hemisphere in summer are much more representative for the composition of the nucleus. Some water is missing that is transported as ice towards the north. An accurate assessment of the nucleus composition requires integration over the whole orbit.

Clear indications of mass transfer on a cometary nucleus surface were first observed during the EPOXI flyby of comet 103P/Hartley 2 (A'Hearn et al. 2011). Centimetre to decimetre sized icy chunks driven by $CO_2$ were observed in large numbers above the active smaller end of the elongated nucleus (Kelley et al. 2013). The area connecting both ends looked smooth. Essentially, here water vapour was observed in the coma. A'Hearn et al. (2011) supposed that this area is covered by the icy grains emitted at the active end of the nucleus. The pronounced inhomogeneity of 103P's coma inferred from the EPOXI flyby does not imply that the composition of the nucleus is inhomogeneous. Large parts of the nuclei of this comet and of 67P are covered by active (water producing) "wet" back fall. Mass transfer on cometary nuclei probably is a frequent phenomenon and may have a strong influence on the evolution of the nucleus. This also complicates the interpretation of coma observations. Observed inhomogeneities do not reflect the bulk composition of the nucleus but are faked by the redistribution of material that is partly desiccated from highly volatile compounds. The Rosetta observations over an almost complete activity cycle provide detailed insight in this process.

It remains a major challenge to accurately determine the fraction of mass transferred to the north. This will only be possible with some reliability when the physical process of activity is understood. Even then the trajectories of large dust particles depend on the homogeneity of the gas streaming from the nucleus. As long as we do not understand the cometary activity using comets for the study of the early solar system will remain ambiguous. The Rosetta mission, set out to understand cometary activity, observed comet 67P over its whole seasonal activity cycle for more than two years. It encountered a very complex world that prevented the orbiter to come close enough to unveil its secrets but posed many new and unanticipated challenges to cometary science, many of which could be solved by a future lander mission visiting this highly variable world.

*Acknowledgements.* OSIRIS was built by a consortium of the Max-Planck-Institut für Sonnensystemforschung, in Göttingen, Germany, CISAS-University of Padova, Italy, the Laboratoire d'Astrophysique de Marseille, France, the Instituto de Astrofisica de Andalucia, CSIC, Granada, Spain, the Research and Scientific Support Department of the European Space Agency, Noordwijk, The Netherlands, the Instituto Nacional de Tecnica Aeroespacial, Madrid, Spain, the Universidad Politecnica de Madrid, Spain, the Department of Physics and Astronomy of Uppsala University, Sweden, and the Institut für Datentechnik und Kommunikationsnetze der Technischen Universität Braunschweig, Germany. The support of the national funding agencies of Germany (DLR), France (CNES), Italy (ASI), Spain (MEC), Sweden (SNSB), and the ESA Technical Directorate is gratefully acknowledged. This project received funding from the European Union's Horizon 2020 research and innovation programme under grant agreement No 686709.

*In memoriam* Mike F. A'Hearn. Much of Rosetta's research has been motivated by the results of the Deep Impact and Epoxy missions that were under Mike's leadership. We lost an outstanding colleague and dear friend.